\documentclass[aps,prb,twocolumn,showpacs,preprintnumbers,amsmath,amssymb,,superscriptaddress,shortbibliography]{revtex4}
\usepackage{graphicx}% Include figure files
\usepackage{dcolumn}% Align table columns on decimal point
\usepackage{bm}% bold math
\usepackage{subfigure}

\begin{document}

%\preprint{APS/123-QED}

\title{Robust surface electronic properties of topological insulators: \\ Bi$_2$Te$_3$ films grown by molecular beam epitaxy}

%\title{
%
%Recovery of surface properties of MBE-grown topological insulator Bi$_2$Te$_3$ after exposure to atmospheric pressure
%
%
%}

% Force line breaks with \\

\author{L. Plucinski}
\email{l.plucinski@fz-juelich.de}
\affiliation{%
Peter Gr\"unberg Institut PGI-6, Forschungszentrum J\"{u}lich, D-52425 J\"{u}lich,
Germany
}%

\author{G. Mussler}
\author{J. Krumrain}
\affiliation{%
Peter Gr\"unberg Institut PGI-9, Forschungszentrum J\"{u}lich, D-52425 J\"{u}lich,
Germany
}%
\affiliation{%
J\"{u}lich Aachen Research Alliance - Fundamentals of Future Information Technologies (JARA-FIT), Germany
}%

\author{A. Herdt}
\affiliation{%
Peter Gr\"unberg Institut PGI-6, Forschungszentrum J\"{u}lich, D-52425 J\"{u}lich,
Germany
}%

\author{S. Suga}
\affiliation{%
Peter Gr\"unberg Institut PGI-6, Forschungszentrum J\"{u}lich, D-52425 J\"{u}lich,
Germany
}%
\affiliation{%
Graduate School of Engineering Science, Osaka University, Toyonaka,
Osaka 560-8531, Japan
}%

\author{D. Gr\"{u}tzmacher}
\affiliation{%
Peter Gr\"unberg Institut PGI-9, Forschungszentrum J\"{u}lich, D-52425 J\"{u}lich,
Germany
}%
\affiliation{%
J\"{u}lich Aachen Research Alliance - Fundamentals of Future Information Technologies (JARA-FIT), Germany
}%

\author{C. M. Schneider}
\affiliation{%
Peter Gr\"unberg Institut PGI-6, Forschungszentrum J\"{u}lich, D-52425 J\"{u}lich,
Germany
}%
\affiliation{%
J\"{u}lich Aachen Research Alliance - Fundamentals of Future Information Technologies (JARA-FIT), Germany
}%

\date{\today}% It is always \today, today,
             %  but any date may be explicitly specified

\begin{abstract}

The surface electronic properties of the important topological insulator Bi$_2$Te$_3$ are shown to be robust under an extended surface preparation procedure which includes exposure to atmosphere and subsequent cleaning and recrystallization by an optimized \emph{in-situ} sputter-anneal procedure under ultra high vacuum conditions. Clear Dirac-cone features are displayed in high-resolution angle-resolved photoemission spectra from the resulting samples, indicating remarkable insensitivity of the topological surface state to cleaning-induced surface roughness.

\end{abstract}

% PACS, the Physics and Astronomy
                             % Classification Scheme.
%\keywords{Suggested keywords}%Use showkeys class option if keyword
                              %display desired
\maketitle

%\section{\label{sec:intro}Introduction}

Recently, a new class of materials, called the topological insulators (TIs), has been predicted theoretically and experimentally observed.\cite{Moore2010} TIs are characterized by odd number of Dirac cones that show a linear energy dispersion similar to relativistic particles. Hence, carriers at the surface of TIs form a two-dimensional electron gas with unparalleled properties, such as extremely high mobilities or dissipationless spin-locked transport, which may lead to new applications in the field of spintronics or quantum computing. Bi$_2$Te$_3$ is a narrow gap semiconductor that has been traditionally investigated as a thermoelectric material, however, very recently the TI behavior has been observed at its surface.\cite{Hsieh2008Nature} To date, the studies of bismuth telluride's TI properties have mainly been carried out on bulk crystals prepared by the melt-growth or self-flux method,\cite{Hsieh2008Nature,Chen2009} which result in heavily doped \emph{n}-type material due to formation of defects. To compensate the n-type doping, the Bi$_2$Te$_3$ crystals are usually heavily doped with Sn or Ca, which, however, strongly degrades transport properties. Looking forward to future device applications, it is desirable to develop growth procedures for intrinsic \emph{undoped} thin films of Bi$_2$Te$_3$, for example by means of molecular beam epitaxy (MBE). Very recently Li \emph{et al.} \cite{Li2010} reported on MBE-growth of high quality single-crystal Bi$_2$Te$_3$ epilayers onto Si(111) substrates that showed the linear energy dispersion as determined by means of \emph{in-situ} angle-resolved photoelectron spectroscopy (ARPES). Implementing such films in device structures will necessarily require some post-growth surface treatment or the formation of interfaces. As the appearance of topologically insulating behavior is tied to electronic states at the boundary of the sample, the question arises how stable these electronic properties are with respect to a subsequent surface modification.

In this letter we report on ARPES measurements of MBE-grown Bi$_2$Te$_3$/Si(111) epilayers which have been exposed to air prior to the measurements. We have observed a remarkable recovery of the TI behavior of the Bi$_2$Te$_3$ surface in the course of a standard cleaning procedure under UHV conditions, which directly proves the insensitivity of the topological surface states to surface roughness and impurities, in agreement with recent scanning tunneling microscopy studies. \cite{Roushan2009,Zhang2009,Park2010,Hanaguri2010}

\begin{figure}
\includegraphics[width=6cm]{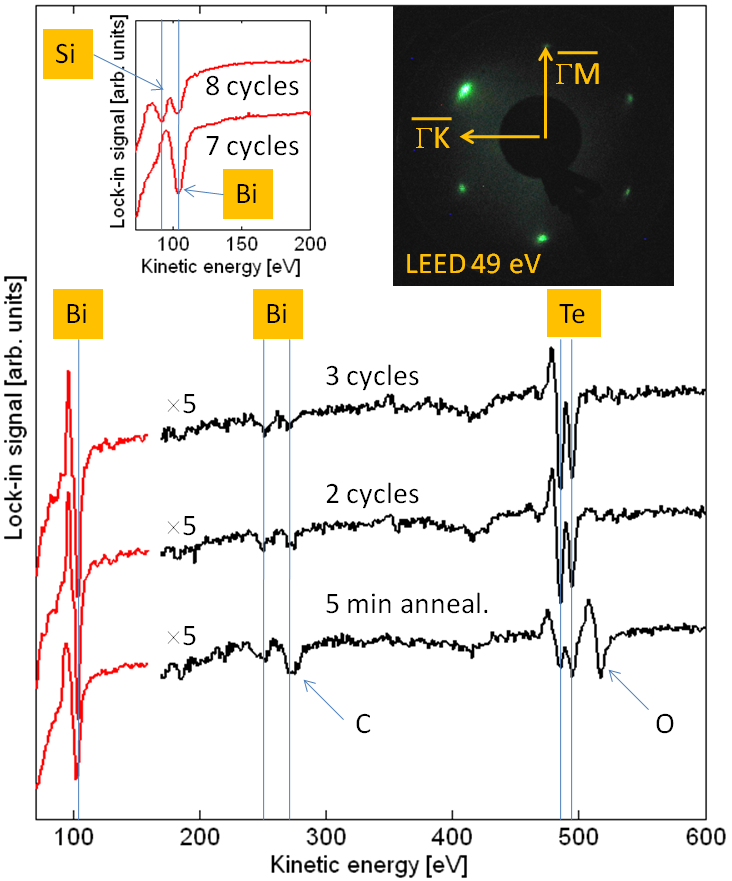}
\caption{\label{fig:auger} Auger spectra showing progress in cleaning of the sample. Each cleaning cycle consisted of 1 min 500 eV Ar-ion sputtering and 5 min annealing at 250$^{\circ}$C. Low energy electron diffraction (LEED) pattern after 2 cleaning cycles is also shown.}
\end{figure}

\begin{figure*}
\includegraphics[width=16cm]{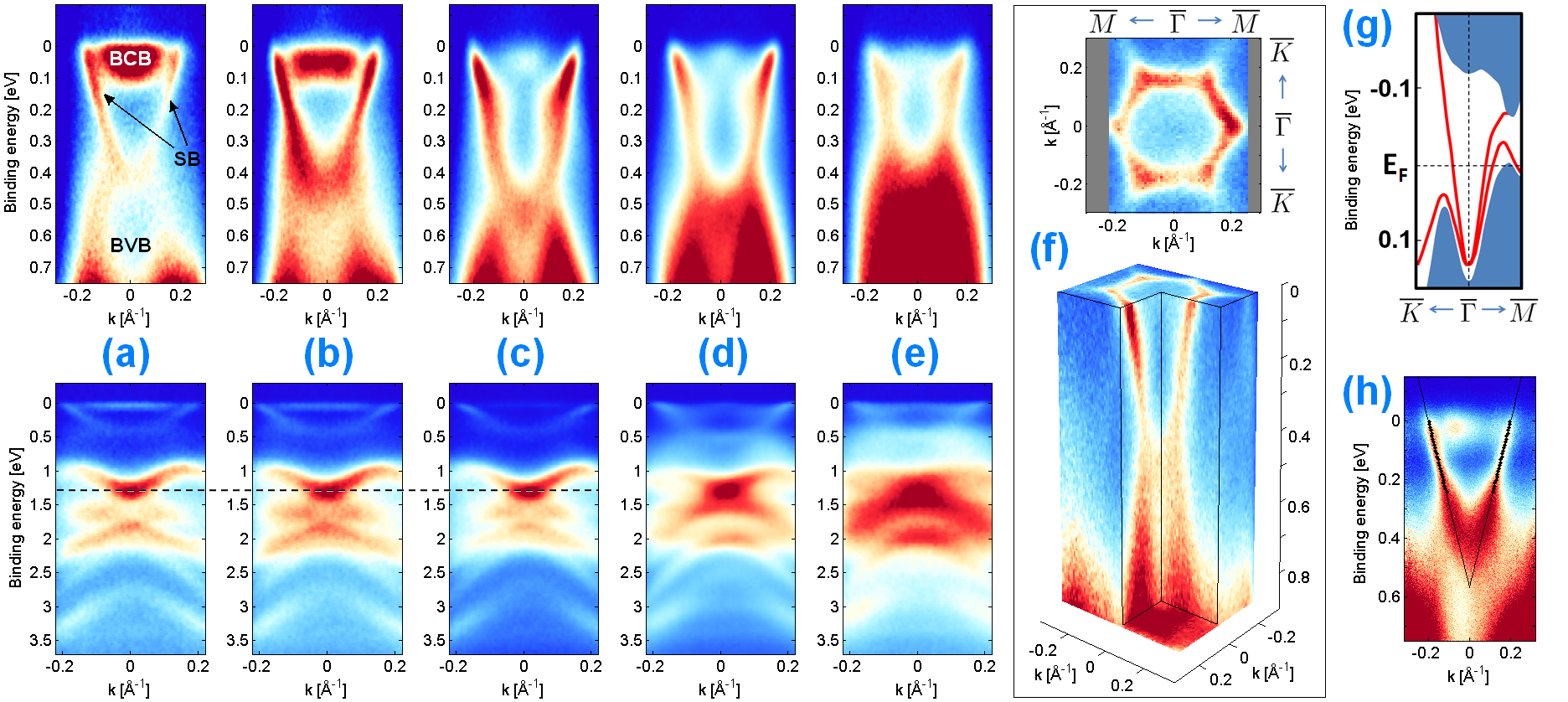}
\caption{\label{fig:arpes} (a)-(e) Selected ARPES maps measured along the $\overline{\Gamma}\overline{K}$ using He I radiation (h$\nu$=21.22 eV) at 140K in the course of cleaning the same Bi$_2$Te$_3$ film several times. (a) 16 QL ($\approx$ 16 nm)  (b) 11 QL (c) 6 QL (d) 3.5 QL (e) 1 QL. Bulk conduction band (BCB), surface state band (SB) and bulk valence band (BVB)  are indicated in panel (a). Dashed line in bottom (a)-(c) panels indicate the unchanged binding energy position of the most pronounced bulk feature. The accuracy of thickness values is $\pm 0.5$ QL ($\pm 0.5$ nm) to account for small instabilities of the sputter gun ion flux and a slightly non-uniform thickness profile. The color scale saturation is separately adjusted in the top and bottom panels. (f) Fermi surface and the three dimensional band structure illustration of the 3.5 QL film measured at 15K. (g) Schematic Bi$_2$Te$_3$ theoretical dispersion as predicted by Zhang \emph{et al.} \cite{Zhang2009theory} (h) 11 QL spectrum along $\overline{\Gamma}\overline{K}$ taken with Kr (10.03 eV) excitation.}
\end{figure*}

%\section{\label{sec:exp}Experiment}

The Bi$_2$Te$_3$ epilayers were grown by means of MBE on \emph{n}-type Si (111) wafer with details of the procedure explained in Ref. \onlinecite{Krumrain2011}. The atomic force microscopy (AFM) images of as-grown films show atomically smooth surfaces with atomic steps of $\approx$ 1 nm height, related to the Bi$_2$Te$_3$ quintuple layer (QL). Pieces of approximately $10\times 10$ mm were cut out of the wafer and transferred into the high resolution ARPES machine.\cite{Suga2010} The temperature of the sample during annealing was controlled by a calibrated infrared pyrometer through the standard UHV viewport. All data shown in this paper were measured using either non-monochromatized He I ($h\nu=21.22$ eV) or monochromatized Kr ($h\nu=10.03$ eV) light.\cite{Suga2010} During the measurements, samples were mounted on the cryostat cooled either by liquid nitrogen or liquid helium, and the resolution of the spectrometer was set to 20 meV for all the presented spectra. The sample has been cleaned by cycles of Ar-ion sputtering\cite{Comment_sputter} at 500 eV and subsequent annealing at 250$^\circ$. The Auger electron spectroscopy (AES) results for one of the films are shown in Fig. \ref{fig:auger}. The appearance of the Si substrate peak in the AES spectra (inset in Fig. \ref{fig:auger}) allowed one to deduce a sputtering rate of $\approx$ 2.5 nm/min. This rate was also confirmed by x-ray reflectivity measurement on other samples taken out of vacuum after a few cleaning cycles. AFM measurements revealed that our cleaning procedure caused an increased surface roughness with an RMS value of $\approx$ 1 nm.

%\section{\label{sec:res}Results and discussion}

Figure \ref{fig:arpes} (a)-(e) shows the set of ARPES $E_B(k)$ maps reflecting the cleaning process of the sample, in which the thickness of the Bi$_2$Te$_3$ decreases with subsequent cleaning cycles. Top panels show the details of the photocurrent near the Fermi level, and bottom panels offer a wider overview, which allows comparison of the bulk and surface features. The most pronounced feature of the spectra is the Dirac cone structure and sharp bulk dispersions, which compare well to the data from cleaved single crystals \cite{Chen2009,Noh2008,Hsieh2009PRL} or films studied \emph{in-situ}.\cite{Li2010} Furthermore, the top panels clearly show the bulk conduction band minimum (CBM) moving upward with decreasing thickness of the film, a process which is accompanied by the increasing valence band spectral weight in comparison to the Dirac cone surface state. Strikingly, this clear movement of $\approx$ 100 meV between panels (a) and (c) is not accompanied by the related shift of the bulk bands (dashed line in these panels indicates unchanged binding energy of the most pronounced bulk feature), therefore it reflects changes in the electronic band structure more complex than just a trivial uniform shift of all bands. For ultrathin films (panel (e)) a clear change in bulk bands is observed, which seems to indicate a collapse of the band structure properties, perhaps because in such thin films, the crystalline properties cannot be reestablished after sputtering.

 Figure \ref{fig:arpes} (f) clearly reveals the complex hexagram shape of the Bi$_2$Te$_3$ Dirac cone Fermi surface dispersion, as seen before by Chen \emph{et al.} \cite{Chen2009} In the Fermi surface image there is no significant spectral weight related to the bulk conduction band at the zone center. In Fig. \ref{fig:arpes}(h) the Kr-spectrum measured on the 11 QL thick film is presented, together with the fit of the Fermi velocity, yielding 2.82 eV$\cdot$\AA (v$_F$=4.28$\cdot 10^5 m/s$), a value similar to the one measured by Chen \emph{et al.} \cite{Chen2009} (2.67 eV$\cdot$\AA, v$_F$=4.05$\cdot 10^5 m/s$) for the $\overline{\Gamma}\overline{K}$ direction.

\begin{figure}
\includegraphics[width=8.5cm]{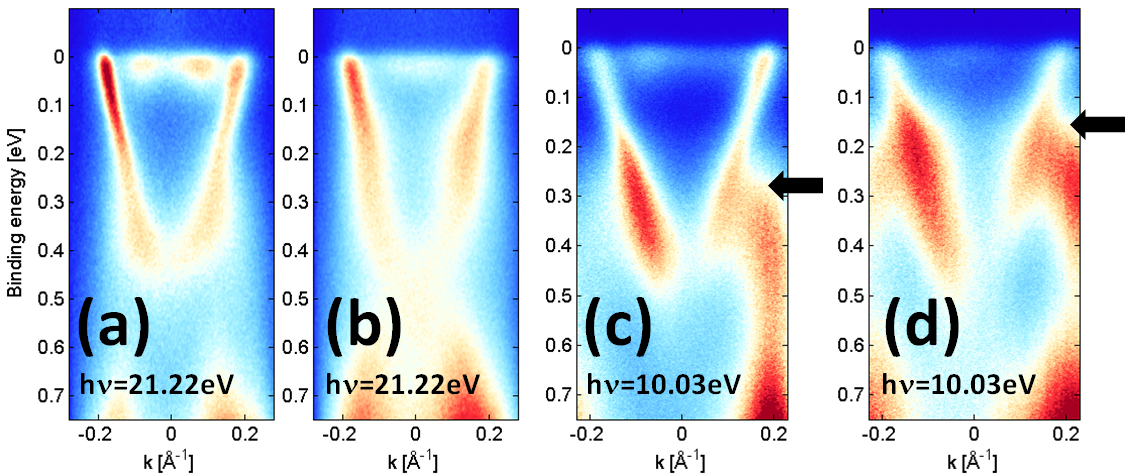}
\caption{\label{fig:slim} Bi$_2$Te$_3$ ARPES maps along $\overline{\Gamma}\overline{K}$ measured at 90K. (a) and (c) Film thickness is 10 QL, (b) and (d) Film thickness is 8.8 QL. (a)-(b): He I excitation, (c)-(d): Kr excitation. The $>$100meV binding energy shift of the significant part of the spectral weight between panels (c) and (d) is indicated by arrows.}
\end{figure}

Figure \ref{fig:slim} shows further ARPES data for two selected thicknesses, measured with both He I and Kr photon sources. In Fig. \ref{fig:slim} (c)-(d) large amount of the additional spectral weight can be seen, which does not belong to the V-like dispersion of the Dirac cone. Between 10 and 8.8 QL thickness this spectral weight exhibits an upward shift of $>$100 meV, which is unexpected taking into account the absence of similar shifts in bulk bands in Fig. \ref{fig:arpes} (a)-(c). Therefore the observed spectral weight might be related to the second dispersive branch of the surface state, as predicted by the theoretical calculations \cite{Zhang2009theory} (see Fig. \ref{fig:arpes} (g)). Small perturbations in the dispersion in panels (a)-(b) can also be found, possibly indicating the same effect, with the difference between the spectra taken at two different photon energies possibly related to the matrix elements effects which may be further enhanced in low dimensional systems.

Our spectra for thicker films (16-11 QL, Fig. \ref{fig:arpes}(a)-(b)) are very similar to the ones of Li \emph{et al.} \cite{Li2010} measured on 2-5 QL films. In their interpretation, the large CBM pocket is related to the thickness of the film, as they have observed a decreasing CBM spectral weight with increased thickness. Since in our data exactly the opposite trend is observed, we postulate that the effect is related to the surface stoichiometry, which in Ref. \onlinecite{Li2010} may be due to the time dependence of the film's growth, and in our case to the selective Ar-ion etching.

Experimental data presented in Figs. \ref{fig:arpes} and \ref{fig:slim}, are representative results obtained on different samples cut from the same wafer. After cleaning procedure all measured samples have shown Dirac cones of the same quality, and the CBM pocket could be removed in all cases.

%\section{\label{sec:sum}Summary}

In conclusion, we have shown that the surface of Bi$_2$Te$_3$/Si(111) thin film can be effectively cleaned under UHV even after exposure to atmospheric conditions for many days. The cleaning process has been controlled by AES and recrystallization of the surface was confirmed by LEED. Photoemission spectra are of similar quality as those obtained on cleaved single crystals or on thin films never exposed to the air. Our results clearly indicate that the coherence of the Dirac-cone topological surface state in Bi$_2$Te$_3$ is, to a large extent, insensitive to the surface roughness induced by the sputter-and-anneal cleaning process, in agreement with other recent studies \cite{Wray2011Iron,Roushan2009,Zhang2009,Park2010,Hanaguri2010} which have shown insensitivity of the TI state to scattering on magnetic and non-magnetic surface impurities. The observed trend of decreasing the CBM spectral weight with increasing sputtering time and decreased film thickness can be very promising from the point of view of further applications and indicates that the robust adjustment of the TI behavior might be achievable by a standard cleaning procedure under the UHV.

We acknowledge stimulating discussions with Caitlin Morgan on the course of editing the manuscript. Thanks are due to NRW Research School "Synchrotron Radiation in Nano- and Bio-Sciences" for financial support.

%\bibliographystyle{apsrev}
%\bibliography{Bi2Te3_biblio}% Produces the bibliography via BibTeX.

\begin{thebibliography}{15}
\bibitem[{Com()}]{Comment_sputter}
\bibinfo{note}{We have used ISE5 cold cathode sputter gun from Omicron GmbH
  which produced approx. 14 $\mu$A sample drain current during the 500 eV
  Ar-ion sputtering.}
\end{thebibliography}

\end{document}